\documentclass[final,aps,prb,twocolumn,superscriptaddress,showpacs]{revtex4-1}

\usepackage{graphicx}
\usepackage{dcolumn}
\usepackage{multirow}
\usepackage{bm,amsmath}
\usepackage{xcolor}
\usepackage{pdfcomment}
\usepackage[normalem]{ulem}
\usepackage{textcomp}

\begin{document}

\newpage

\title{Electronic characteristics of ultrathin SrRuO$_{3}$ films \\and their relationship with the metal--insulator transition}

\author{Subeen~Pang}
\affiliation{Department of Materials Science and Engineering and Research Institute of Advanced Materials,
             Seoul National University, Seoul 08826, Republic of Korea}

\author{Yoonkoo~Kim}
\affiliation{Department of Materials Science and Engineering and Research Institute of Advanced Materials,
             Seoul National University, Seoul 08826, Republic of Korea}

\author{Yeong~Jae~Shin}
\affiliation{Center for Correlated Electron Systems, Institute for Basic Science (IBS), Seoul 08826, Republic of Korea}
\affiliation{Department of Physics and Astronomy,
				Seoul National University, Seoul 08826, Republic of Korea}

\author{Byungmin~Sohn}
\affiliation{Center for Correlated Electron Systems, Institute for Basic Science (IBS), Seoul 08826, Republic of Korea}
\affiliation{Department of Physics and Astronomy,
				Seoul National University, Seoul 08826, Republic of Korea}

\author{Seung‐Yong~Lee}
\affiliation{Department of Materials Science and Engineering and Research Institute of Advanced Materials,
             Seoul National University, Seoul 08826, Republic of Korea}

\author{Tae~Won~Noh}
\affiliation{Center for Correlated Electron Systems, Institute for Basic Science (IBS), Seoul 08826, Republic of Korea}
\affiliation{Department of Physics and Astronomy,
				Seoul National University, Seoul 08826, Republic of Korea}

\author{Miyoung~Kim}
\email[Corresponding author. E-mail: ]{mkim@snu.ac.kr}
\affiliation{Department of Materials Science and Engineering and Research Institute of Advanced Materials,
             Seoul National University, Seoul 08826, Republic of Korea}

\begin{abstract}

SrRuO$_3$ (SRO) films are known to exhibit insulating behavior as their thickness approaches four unit cells. We employ electron energy-loss (EEL) spectroscopy to probe the spatially resolved electronic structures of both insulating and conducting SRO to correlate them with the metal--insulator transition (MIT). Importantly, the central layer of the ultrathin insulating film exhibits distinct features from the metallic SRO. Moreover, EEL near-edge spectra adjacent to the SrTiO$_3$ (STO) substrate or to the capping layer are remarkably similar to those of STO. The site-projected density of states based on density functional theory (DFT) partially reflects the characteristics of the spectra of these layers. These results may provide important information on the possible influence of STO on the electronic states of ultrathin SRO.

\end{abstract}



\maketitle

%

\section{Introduction}
\label{Introduction} 

\begin{figure}[t]%
\begin{center}
\includegraphics*[width=0.9\linewidth]{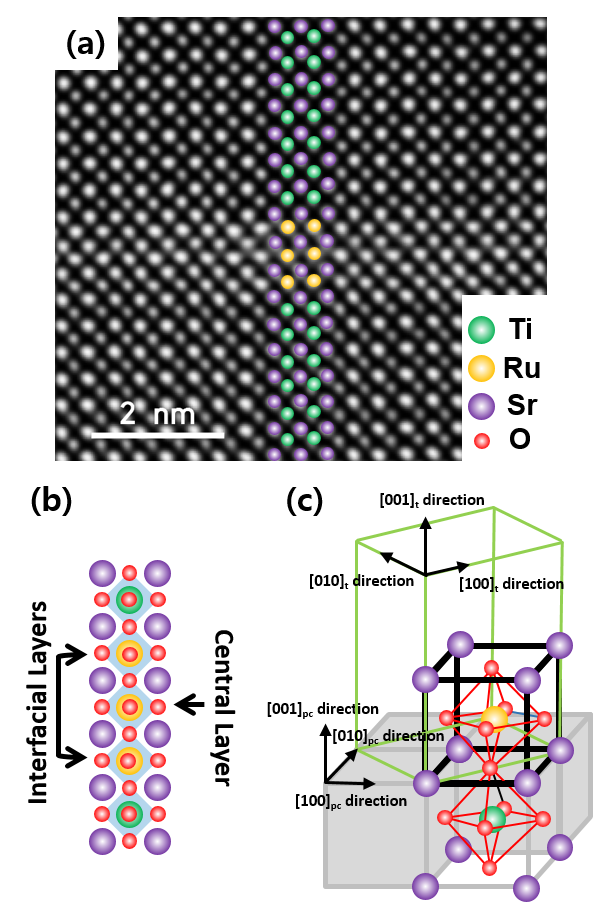}
\end{center}
\caption{%
	(a) High-angle annular dark-field (HAADF) image of capped SrRuO$_3$ (SRO) with three-unit-cell thickness. (b) Schematic diagram of capped SRO. (c) Pseudo-cubic (pc) and tetragonal (t) lattice structure of ultrathin SRO grown on a SrTiO$_3$ (STO) substrate without capping. Purple, yellow, green, and red circles indicate Sr, Ru, Ti, and O atoms, respectively, as indicated in (a).
}
\label{structure}
\end{figure}

SrRuO$_3$ (SRO) is a ferromagnetic metal oxide with a Curie temperature $T_c$ of 160 K \cite{j1}. It is often used as a gate electrode due to its high conductivity and ease of epitaxial growth \cite{j2}. SRO has attracted considerable attention because of its intriguing electronic behaviors, for example, SRO loses its itinerant ferromagnetism as the thickness approaches approximately four unit cells \cite{j3,j4,j5}. Such behavior, referred to as the metal-insulator transition (MIT) of SRO, is a stark deviation from the property of bulk SRO which is known to be only weakly correlated \cite{j6}. There have been many theoretical efforts to explain the origin of the MIT. \\
  \indent In density functional theory (DFT) studies, the on-site Coulomb interaction parameter, $U$, has been used to model electronic correlations of $d$-orbitals.  However, DFT+$U$ calculations of ultrathin SRO were inconsistent with experimental results. For instance, SRO remained metallic regardless of unrealistic $U$ values \cite{j23} and extreme (one-unit-cell) thickness. \cite{j24}  Hence, in addition to $U$, extrinsic factors such as surface relaxation, in-plane strain, and disorder have been suggested as possible origins of the MIT. For instance, the effective Coulomb potential of about 2--3 eV in the presence of high surface relaxation \cite{j20,j21,j22} or DFT+$U$ under large tensile strain produced the insulating phase \cite{j9}. However, experimentally, SRO/ultrathin SRO/STO under compressive strain without surface relaxation or reconstruction also exhibited insulating behavior \cite{j13}. Hence, additional DFT+$U$ studies are required to demonstrate a clear description of the insulating SRO. \\
  \indent The origin of such theoretical difficulty arises from two main factors. First, the physics of ultrathin SRO cannot be described precisely by only a few parameters, such as $U$ \cite{j23} and in-plane strain exerted by STO \cite{j9,j24}, and delicate structural alteration may result in drastic changes in physical properties. For these reasons, Hund's coupling \cite{j33}, dynamical correlation \cite{j23} and dimensionality reduction \cite{j3,j24}, as seen in the MIT of SrVO$_3$ \cite{j34}, have been proposed as possible causes of the MIT as well. Second, the electronic structure is expected to differ layer by layer within ultrathin systems, regardless of surface relaxation \cite{j32}. Such an exacting nature of the ultrathin system can exhibit unreported electronic or magnetic behaviors (e.g., comparing Ref. 8 with 11). Hence, to fully understand the MIT of SRO, we first need to carefully analyze the electronic structure of each atomic layer of ultrathin SRO.\\
  \indent  Experimentally, SRO exhibited electronic anomalies that could not be explained in terms of $U$ \cite{j49,j50,j51}. Furthermore, although photonic spectroscopy studies have reported a vestige of the lower Hubbard band \cite{j15,j16} and hard-gap originating from spectral incoherency in  thin-film SRO \cite{j10}, some studies have criticized that the highly correlated spectra of surface SRO would contaminate and thus exaggerate such incoherency \cite{j17,j18}. These results imply that the MIT in SRO cannot be explained merely with Mott-Hubbard physics, which is consistent with what has been predicted in theoretical studies. \\
  \indent On the other hand, the physics of ultrathin SRO significantly depended on substrates in experiments. For instance, SrTiO$_3$ (STO) substrates exert compressive in-plane strain, making rotations and tilts of oxygen octahedra of SRO energetically unfavorable. As a result, the lattice system of ultrathin SRO underwent orthorhombic-to-tetragonal phase transformation (Figure \ref{structure}(c)) and magnetization was significantly suppressed \cite{j28,j29,j30}. Furthermore, the tetragonality increased as the thickness of SRO film decreased \cite{j29}. Hence, we need to scrutinize not only the layer-by-layer dependence of the electronic states but also the effects of the tetragonality and substrates in ultrathin SRO.  \\
  \indent In this letter, we report electron energy-loss (EEL) spectroscopy of SRO with three-unit-cell (capped insulator) and 24 nm (metal) thicknesses near the O-$K$ edge ($1s\rightarrow2p$ transition). Interestingly, we find that the central layer of the insulating SRO exhibits distinct features from metallic and interfacial SRO. To identify whether the features originate from in-plane strain and the corresponding strong tetragonality, we perform DFT calculations of the STO/SRO/STO superlattice with highly suppressed tilts and rotations of oxygen octahedra \cite{j48}, and compare them with SRO bulk with a tetragonal crystal-field. By analyzing the characteristics of the spectra and computational results, we provide comments on the possible origins of the MIT.

\section{Experimental}
\label{Theoretical} 

SRO films were grown on a (001) TiO$_2$-terminated STO substrate. On the substrate, we deposited SRO films using pulsed laser deposition with an oxygen pressure of 0.1 Torr and a laser fluence of 1.5 J/cm$^2$ at 700 \textdegree{}C. The growth rate of SRO films was approximately 0.013 nm/s. A focused ion beam was used to prepare specimens for transmission electron microscopy (TEM) analysis. A JEOL-ARM200F scanning TEM (STEM) provided high-angle annular dark-field (HAADF) images and EEL spectra near the O-$K$ edge. The scanning rate of the STEM-EELS detector was 0.1 s/pixel. The electrical resistivity of SRO films was measured using the standard four-probe method.\\
\indent To calculate the density of states (DOS) of the (STO)$_3$/(SRO)$_3$/(STO)$_3$ superlattice (S3) and tetragonally elongated SRO (ST), we adopted computational procedures similar to those used in Ref. 19. The calculations were performed using the plane-wave basis set and the projector-augmented wave method implemented in the Vienna ab initio simulation package \cite{j45}. We used the generalized gradient approximation with a PBEsol functional \cite{j46}. Starting from ferromagnetic configuration, we chose a weak correlation $U_{\text{eff}}$ = 1 eV, which is suitable to approximate the experimental results \cite{j22,j23}. A plane-wave energy cutoff of 500 eV was used with Monkhorst-Pack mesh $k$-point sampling of 21 $\times$ 21 $\times$ 1 for S3, and 21 $\times$ 21 $\times$ 21 for ST. The samplings were checked up to 41 $\times$ 41 $\times$ 1 and 61 $\times$ 61 $\times$ 61, respectively. From the Poisson effect, the pseudo-cubic out-of-plane parameter of SRO on a STO substrate was estimated to be approximately 3.9635 {\AA} \cite{j25,j27,j36,j37,j38}. Hence, for S3 and ST, we fixed the pseudo-cubic in-plane lattice parameter of SRO to be 3.905 {\AA}, which is a lattice parameter of cubic STO, and the out-of-plane parameter to be 3.9635 {\AA}.

\begin{figure}[t]%
\begin{center}
\includegraphics*[width=0.7\linewidth]{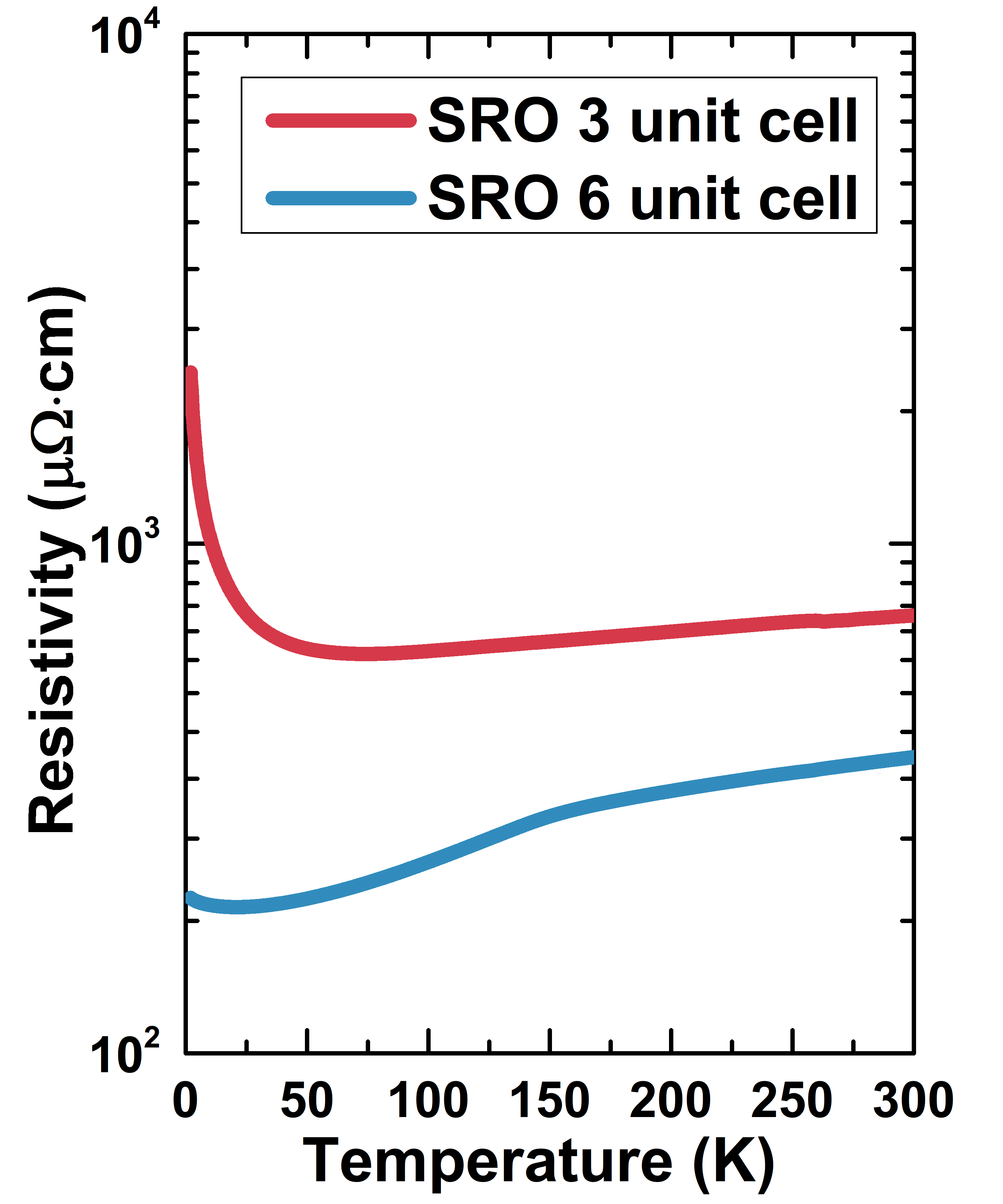}
\end{center}
\caption{%
	Temperature-dependent resistivity of SrRuO$_3$ (SRO) specimens with thicknesses of six and three unit cells, measured using the standard four-probe method.
}
\label{res}
\end{figure}

\section{Results and Discussion}
In Figure \ref{structure}(a), SRO with a three-unit-cell thickness is capped by STO. In this environment, large surface relaxation cannot occur. Figure \ref{res} presents the resistivity data of the capped three- and six-unit-cell SRO specimens, which exhibited similar tendencies to those reported previously \cite{j3,j10,j39}. The resistivity of SRO with a six-unit-cell thickness was proportional to temperature, meaning that it was metallic. In addition, at approximately 160 K, a slope change occurred, which is a typical behavior of ferromagnetic materials near the Curie temperature. On the other hand, at a three-unit-cell thickness, SRO lost both of its ferromagnetic and metallic behaviors (SRO MIT). Hence, we reconfirmed that surface reconstruction is not a generic origin of the MIT.

\begin{figure}[t]%
\begin{center}
\includegraphics*[width=0.8\linewidth]{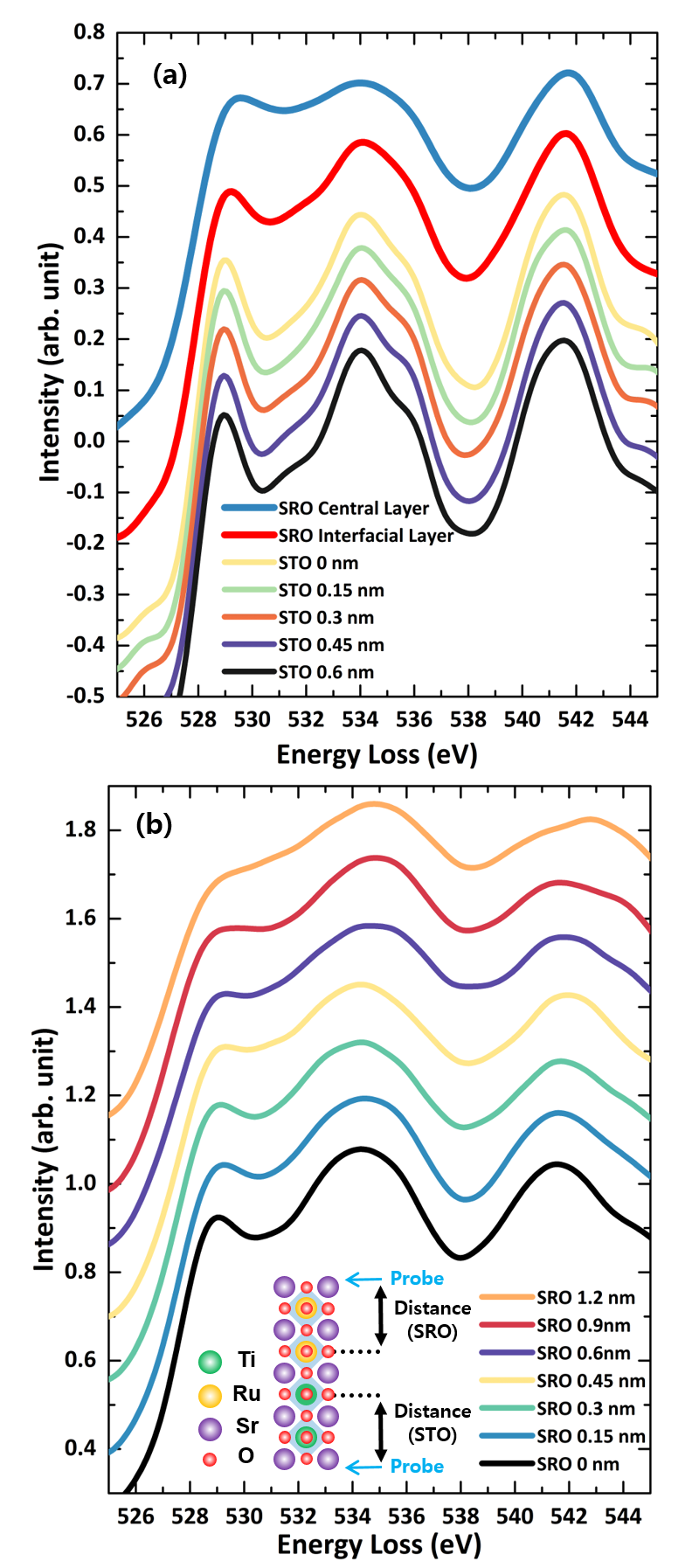}
\end{center}
\caption{(a) Electron energy-loss (EEL) spectra of capped SrRuO$_3$ (SRO) with three-unit-cell thickness near O-$K$ edge. (b) EEL spectra of SRO (24 nm)/SrTiO$_3$ (STO) near O-$K$ edge. Distances indicated in the legend correspond to the separation between the SRO/STO interface and the position of the EEL probe as indicated in the inset of (b).}
\label{3uc}
\end{figure}


To scrutinize the electronic structure of SRO under the MIT, we measured EEL spectra of both STO and SRO regions in our three-unit-cell SRO specimen (Figure \ref{3uc}(a)). Central and interfacial layers in the three-unit-cell SRO indicate specific regions shown in Figure \ref{structure}(b). The overall shapes of EEL spectra of STO were in agreement with a previous report \cite{j40}, showing a three-peak feature with peaks located at approximately 529, 534, and 542 eV. Based on a previous study \cite{j41}, the onset peak near 529 eV in the STO region is associated with the $t_{2g}$ states of Ti, and the second peak at 534 eV is related to the $e_g$ states.

EEL spectra of SRO also displayed the three-peak feature. Considering that the $4d$ orbitals of Ru and $2p$ orbitals of O are highly hybridized \cite{j35}, the first two peaks are designated as $t_{2g}$- and $e_g$-related states. Note that the O-$K$ spectrum of SRO adjacent to STO is very similar to that of STO, in both metallic and insulating SRO films. More importantly, the O-$K$ edge at the central layer of insulating SRO is clearly distinct from the others: the central layer of metallic SRO, two-unit-cell off (same distance from the interface) the interface of metallic SRO, and insulating SRO near the interface. For instance, the intensity of $t_{2g}$-related states significantly increases at the central SRO in Figure \ref{3uc}.

At the SRO/STO interface, SRO with the thickness of 24 nm (Figure \ref{3uc}(b)) displayed the STO-overlapped three-peak features, analogous to Figure \ref{3uc}(a); however, as it was far from the interface, intermediate signals emerged between 529 and 534 eV. Furthermore, the peak of the $t_{2g}$-related states became ambiguous in the metallic region, which is distinguishable from the features of the central spectra of the three-unit-cell SRO. These results show that we cannot ignore the influences of the STO substrates near the interfaces regardless of the thickness of the SRO film and the electronic states of the central layer of the insulating SRO cannot be explained merely by the interfacial effects.

The in-plane strain and strong (tetragonal) crystal-field significantly modify the electronic states of ultrathin SRO \cite{j8,j9,j20}. The STO substrates suppress the rotations and tilts of the oxygen octahedra of SRO \cite{j28,j29,j30,j48}. The resultant lattice system is tetragonal, as shown in Figure \ref{structure}(c). In this manner, unlike other uncapped SRO thin films, the tetragonal lattice system induced by the suppression of rotations along in-plane axes would be further suppressed in our capped SRO. Hence, the strain and strong tetragonality can be correlated with the intriguing behaviors of the spectra. To theoretically analyze the spectra and effects of the strain and tetragonality, we used DFT to obtain the projected density of states of our system and adopted compressively strained SRO structures (S3 and ST) without the rotations and tilts of the oxygen octahedra.

\begin{figure}[t]%
  \includegraphics*[width=\linewidth]{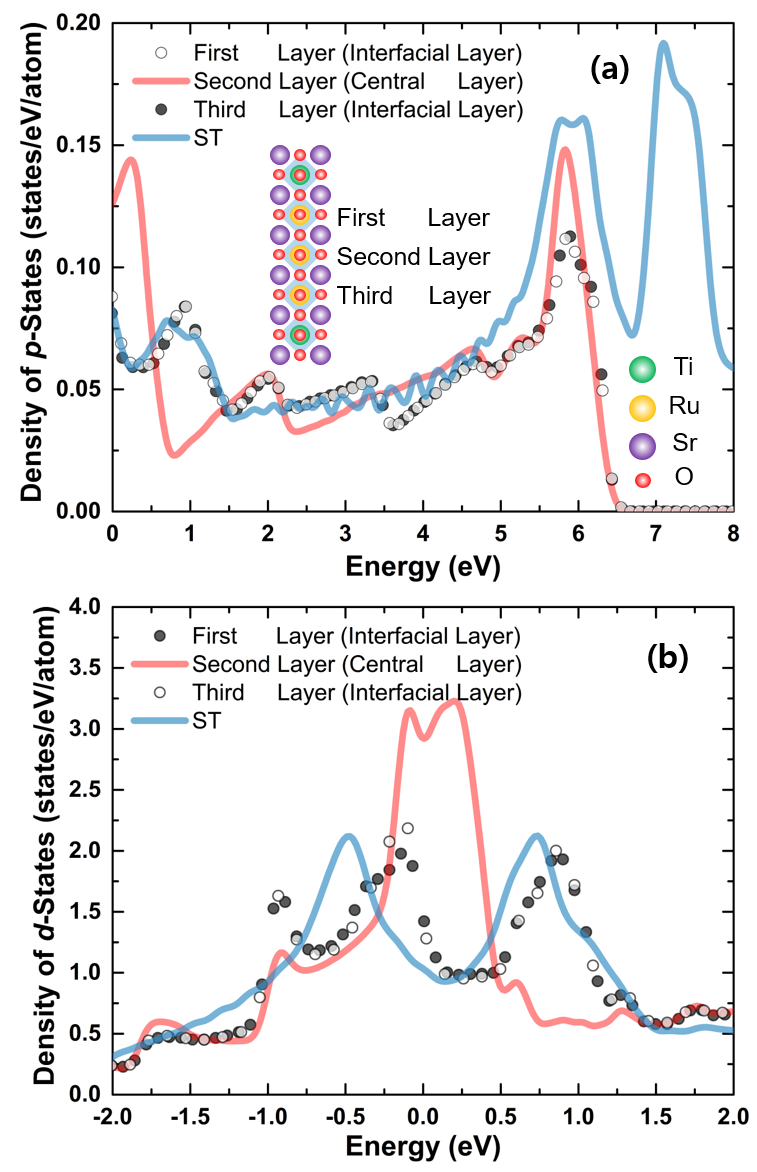}%

  \caption{(a) Projected density of $p$-states of oxygen atoms and (b) projected density of $d$-states of Ru atoms in ST and S3 calculated by density functional theory with $U_{\text{eff}}=1$ eV. The Fermi level is at 0 eV. Positions of the first, second, and third layers are shown in (a). Original densities of states were convoluted by a Gaussian function with a width of 0.1 eV.}
    \label{dos}
\end{figure}

Figure \ref{dos}(a) shows the projected densities of the $p$-states of oxygen atoms in S3; as expected from the EEL spectra, clear differences were observed among the central and interfacial layers. Particularly, the central layer had a more pronounced DOS at the Fermi level compared to the interfacial layers. At approximately 6 eV above the Fermi level, we can see $e_g$-related DOS. However, none of the layers displayed insulating behavior (Figure \ref{dos}), which is inconsistent with the experimental results.

In the EEL spectra, the interfacial regions were significantly influenced by STO irrespective of the thickness of SRO (Figure \ref{3uc}(a) and (b)), while the central layer retained the originality with respect to both the interfacial and bulk regions. In other words, the electronic states of the capped SRO are highly overlapped with those of STO, and the originality of the central layer is probably due to the relatively weak overlap compared to the interfacial regions. Hence, the effective crystal-field splitting may not be the critical component of the system. Rather, the overlap with STO is important. In fact, although the calculation results for S3 were not perfectly consistent with the experimental results, they also showed the potential influences of the hybridization with STO. The electronic states of the central layer of S3 differed from those of ST, i.e., SRO under a high tetragonal crystal-field and without STO substrates (Figure \ref{dos}). This indirectly suggests that hybridization cannot be overlooked.

On the other hand, it is also possible that the STO substrates induce dimensionality reduction (i.e., abrupt truncation of wave function of SRO). In this case, van Hove singularity results in distinct electronic states irrespective of the hybridization with STO \cite{j3,j24}. However, we find no signs of low dimensionality in either the experimental or the computational results. For instance, the projected DOS of $d$-states differed significantly layer by layer (Figure \ref{dos}(b)). Furthermore, as mentioned earlier, the interfacial regions were highly overlapped with STO, meaning that the wave function may not be truncated at the SRO/STO interface.
Lastly, although we adopted a lower correlation ($U_{\text{eff}}$) compared to some studies (e.g., Ref. 17, 18, 21), merely increasing $U_{\text{eff}}$ would not guarantee the successful reproduction of real SRO \cite{j42}. To confirm whether a high correlation results in the insulating state, we produced the DOS of S3 with an unrealistic correlation (i.e., $U_{\text{eff}}=6$ eV). However, it was neither insulating nor more consistent with the experimental results than S3 with $U_{\text{eff}}=1$ eV (see Supporting Information). In other words, simply adopting the high localization and strong tetragonality would not lead to the display of the MIT of SRO. Hence, it is possible that dynamic correlation is important in this system \cite{j23}. With such dynamic effects, the relationship between the hybridization with STO and the formation of the insulating SRO should be checked.

\section{Summary and Conclusion}
We fabricated a STO/SRO (three-unit-cell)/STO system to identify the electronic characteristics of ultrathin SRO. HAADF-STEM images showed an atomistically sharp interface and EEL spectra revealed that the electronic state of the central SRO differs from that of interfacial and bulk SRO even at a three-unit-cell thickness. Particularly, the $t_{2g}$-related states of central SRO are suspected to represent distinct physics. To theoretically analyze the EEL spectra, we performed DFT calculations. However, even if we highly constrained the rotational degrees of freedom and artificially maintained the tetragonality of the superlattice, the calculation did not sufficiently reflect the experimental results. Based on our theoretical and experimental results, we expect that consideration of extra degrees of freedom, other than the effective crystal-field, high localization, and van Hove singularities, may be required to explain the original features of the central layer.

\newpage

\section{Supporting Information}
\begin{figure}[ht]%
  \includegraphics*[width=\linewidth]{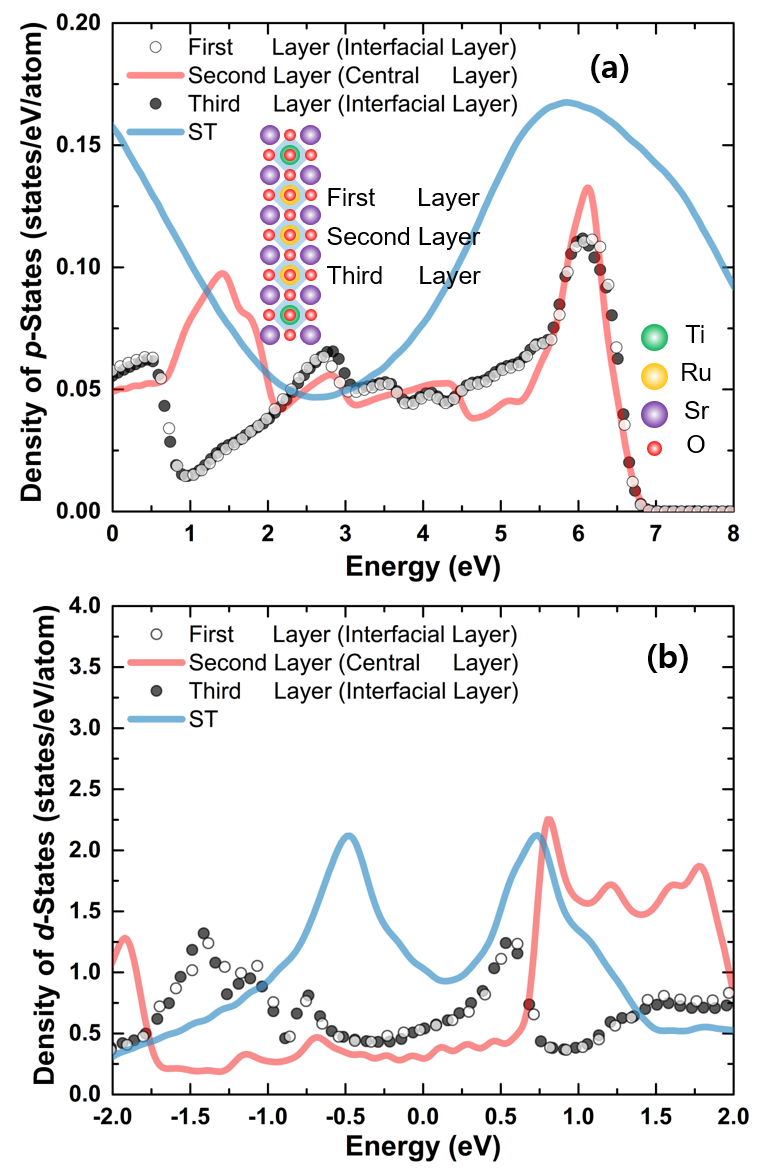}%

  \caption{(a) Projected density of $p$-states of oxygen atoms and (b) projected density of $d$-states of Ru atoms in ST and S3 calculated by density functional theory with $U_{\text{eff}}=6$ eV. The Fermi level is at 0 eV. Positions of the first, second, and third layers are shown in (a). Original densities of states were convoluted by a Gaussian function with a width of 0.1 eV.}
    \label{u6}
\end{figure}

To confirm whether an unrealistically high $U$ and strong tetragonality produces the insulating phase, we calculated the DOS of S3 with $U_{\text{eff}}=6$ eV (Figure \ref{u6}). Although the DOS of Ru $d$-states near the Fermi level was significantly decreased compared to Figure \ref{dos} (b), our system did not exhibit the insulating state. Furthermore, at the Fermi level, the DOS of O $p$-states of the central layer was lowered, which is inconsistent with our experimental results (Figure \ref{3uc} (a)).

\end{document}